\documentclass[apjl]{emulateapj}

\usepackage{todonotes}
\usepackage{color}
\usepackage{multirow}
\usepackage[colorlinks,urlcolor=blue,citecolor=blue,linkcolor=blue]{hyperref}
\usepackage{natbib}
\usepackage{graphics}
\usepackage{subfigure}

\newcommand{\btxt}[1]{{#1}}

\slugcomment{To appear in ApJ Letters}

\shorttitle{The compressed wind of $\eta$~Car~A}
\shortauthors{Teodoro et al.}

\begin{document}

\title{Detection of the compressed primary stellar wind in $\eta$~Carinae$^\star$}

\author{M.~Teodoro\altaffilmark{1,2}}
\author{T.~I.~Madura\altaffilmark{1,3}}
\author{T.~R.~Gull\altaffilmark{1}}
\author{M.~F.~Corcoran\altaffilmark{4,5}}
\author{K.~Hamaguchi\altaffilmark{4,6}}

\affil{$^1$Astrophysics Science Division, Code 667, NASA Goddard Space Flight Center, Greenbelt, MD 20771, USA}

\altaffiltext{2}{CNPq/Science without Borders Fellow.}
\altaffiltext{3}{NASA Postdoctoral Program Fellow.}
\altaffiltext{4}{CRESST and X–ray Astrophysics Laboratory, Code 662, NASA Goddard Space Flight Center, Greenbelt, MD 20771, USA}
\altaffiltext{5}{Universities Space Research Association, 10211 Wincopin Circle, Suite 500 Columbia, MD 21044, USA}
\altaffiltext{6}{Department of Physics, University of Maryland, Baltimore County, 1000 Hilltop Circle, Baltimore, MD 21250, USA.}

\altaffiltext{$\star$}{Based on observations made with the NASA/ESA Hubble Space Telescope, obtained at the Space Telescope Science Institute, which is operated by the Association of Universities for Research in Astronomy, Inc., under NASA contract NAS 5-26555. These observations are associated with program numbers 12013, 12508, and 12750. Support for program numbers 12013, 12508, and 12750 was provided by NASA through a grant from the Space Telescope Science Institute, which is operated by the Association of Universities for Research in Astronomy, Inc., under NASA contract NAS 5-26555.}

\email{mairan.teodoro@nasa.gov}

\begin{abstract}
A series of three \textit{HST/STIS} spectroscopic mappings, spaced approximately one year apart, reveal three partial arcs in [\ion{Fe}{2}] and [\ion{Ni}{2}] emissions moving outward from $\eta$~Carinae. We identify these arcs with the shell-like structures, seen in the 3D hydrodynamical simulations, formed by compression of the primary wind by the secondary wind during periastron passages.
\end{abstract}

\keywords{Stars: individual ($\eta$~Carinae) --- stars: massive --- binaries: general}

\section{Introduction}

$\eta$~Carinae (hereafter $\eta$~Car) is now recognized to be a massive binary system in a highly eccentric orbit ($e>0.9$). The primary star, $\eta$~Car~A, is in the luminous blue variable (LBV) phase, with a luminosity of $10^{6.7}~L_\sun$ and mass of about $120~M_\sun$ \citep{1997ARA&A..35....1D}. It is surrounded by a massive stellar wind \citep[$\dot{M}\sim10^{-3}$~M$_\sun$\,yr$^{-1}$, $v_\infty\approx420$~km\,s$^{-1}$;][]{2012MNRAS.423.1623G} that interacts with a companion, $\eta$~Car~B, that has not been detected directly yet. The physical parameters of $\eta$~Car~B are, thus, inferred from X-ray \citep[$\dot{M}\sim10^{-5}$~M$_\sun$\,yr$^{-1}$, $v_\infty\approx3000$~km\,s$^{-1}$;][]{2002A&A...383..636P} and photoionization modeling \citep[$\sim10^5~L_\sun$, $34000$~K$<T_{\rm eff}<\btxt{39000}$~K;][]{2005ApJ...624..973V,2008MNRAS.387..564T,2010ApJ...710..729M}.

%Three-dimensional (3D) hydrodynamic simulations have greatly contributed to our understanding of the interaction of massive winds in $\eta$~Car \citep[][T. Madura et al. 2013, in preparation]{Parkin:2008dv,Parkin:2009ei,Parkin:2010hj,2012MNRAS.420.2064M}. When comparing the 3D simulations with the observations \citep{2009MNRAS.396.1308G,2011ApJ...743L...3G}, the conclusion is that 
The orbital orientation is such that the orbital plane has an inclination of $i\approx140\arcdeg$, with the semi-major axis pointing along position angle (P.\,A.\footnote{P.\,A. is defined as the angle, in the plane of the sky, eastward of North.}) $314\arcdeg$ and longitude of periastron $\omega\approx240\arcdeg$ \citep{2009MNRAS.396.1308G,2009MNRAS.394.1758P,2011ApJ...743L...3G,2012MNRAS.420.2064M}. In this configuration, $\eta$~Car~B is in front of $\eta$~Car~A at apastron, and the orbital angular momentum vector is parallel to the symmetry axis of the bipolar Homunculus nebula that surrounds the system.

Projected on the sky, \btxt{at apastron,} the secondary star directly illuminates and photo-ionizes the circumstellar material within 1200 AU of the binary system primarily to the northwest, where the material is primarily approaching the observer. The companion star spends most of the orbit near apastron, so that the undisturbed wind from $\eta$~Car~A is located on the far side of the system, red-shifted as seen from Earth, and largely unaffected by the ionizing radiation from $\eta$~Car~B.

During periastron passage, $\eta$~Car~B approaches within $1-2$ AU of $\eta$~Car~A, leading to rapid changes in the shape of the wind-wind interaction surface and confinement of the far-UV flux from $\eta$~Car~B. Hydrodynamical simulations \citep[][T. Madura et al. 2013, in preparation]{2011ApJ...726..105P,2012MNRAS.420.2064M} show that, after each periastron passage, a highly-distorted volume of hot, low density secondary wind pushes outward into the slow, high-density primary wind, leading to the formation of a thin, high density wall surrounding the lower density, trapped wind of $\eta$~Car~B. This dense wall is accelerated to velocities somewhat higher than the terminal velocity of $\eta$~Car~A and expands both in the orbital plane and perpendicular to it, creating a thin, high-density sheet of trapped primary wind material.

%while expanding faster in the orbital plane, expands radially above and below the orbital plane creating a thin, high density, distorted sheet.

Using spectral imaging maps from the Space Telescope Imaging Spectrograph (\textit{STIS}), on board the Hubble Space Telescope (\textit{HST}), we identify structures which are consistent with these walls of compressed primary wind material. We identify three partial arcs formed by the close passage of $\eta$~Car~B around the primary star over the last 3 orbital cycles, and using the \textit{STIS} maps, we determine their space velocity and age.

\section{Observations, data reduction, and analysis}

%\subsection{Observations}

Observations using \textit{HST/STIS} were accomplished through three guest observer programs, 12013 (2010 October 26), 12508 (2011 November 20), and 12750 (2012 October 18), using the G430M and G750M gratings, centered at 4706\AA~and 7283\AA, respectively, in combination with the $52\arcsec\times1\arcsec$~aperture. An extended region, covering $1\arcsec$ to $2\arcsec\times6.4\arcsec$, centered upon $\eta$~Car, was mapped using $0.05\arcsec$~spacing offsets at P.\,A. constrained by the spacecraft. Exposures were approximately 30 seconds using CRSPLIT=2. An additional 7 position sub-map, also centered on $\eta$~Car, employed shorter exposures to allow for potential saturation of continuum on the central positions.

\begin{figure}[!t]
\centering
\resizebox{\columnwidth}{!}{\includegraphics{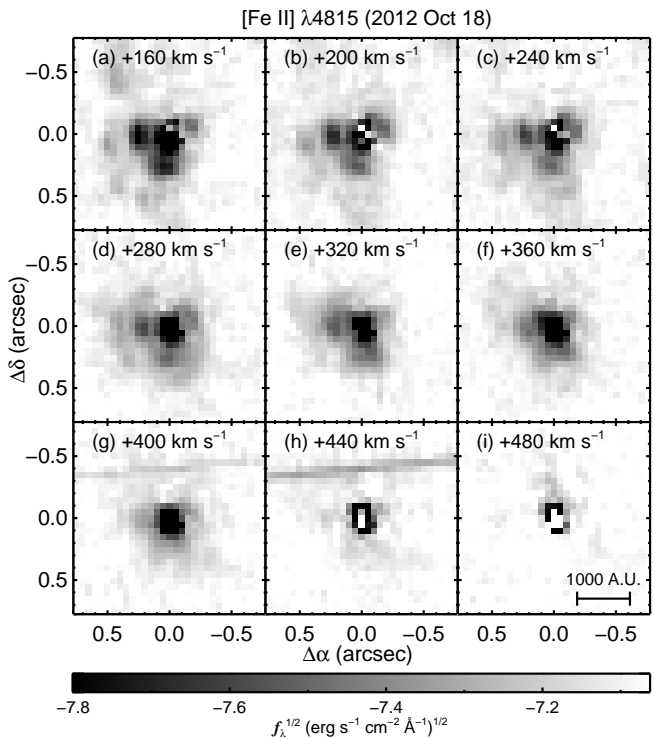}} \caption{Iso-velocity images of [\ion{Fe}{2}]~$\lambda4815$ for the observations of 2012 October 18. The velocities are indicated in each panel. The linear structures that appear in panels ($g$) and ($h$) are the result of a bad CCD column. The physical scale is based on an adopted distance of 2350 pc \citep{2006ApJ...644.1151S}. We note that there are no extended structures at Doppler velocities higher than $+440$~km\,s$^{-1}$.\label{fig:panel}}
\end{figure}

Data were reduced with \textit{STIS} GTO \href{http://www.exelisvis.com/ProductsServices/IDL.aspx}{\sc{idl}}-based\footnote{{\sc idl} is a trademark of Exelis Visual Information Solutions, Inc.} tools similar to the standard pipeline reduction tools \citep{Bostroem:2011wg}, but with improved ability to spatially align the spectra to sub-pixel accuracy. For each emission line of interest, continuum levels were fitted for each spatial position along the aperture and a re-sampled data cube was generated, with the coordinates of right ascension and declination at $0.05\arcsec$~spacing, and Doppler velocity at 20~km\,s$^{-1}$ intervals.

Since we are interested in structures formed in the primary wind around periastron, here we focus only on the following low-ionization lines: [\ion{Fe}{2}] ($\lambda_0=4729.39$, 4776.05, 4815.88, 7157.13)\footnote{All rest wavelengths, $\lambda_0$, are in \AA, in vacuum \citep{2012A&A...540A.133Z}.} and [\ion{Ni}{2}] ($\lambda_0=7413.65$). Observations from program 12013 (2010 October 26) did not include maps of [\ion{Fe}{2}]~$\lambda7157.13$ and [\ion{Ni}{2}]~$\lambda7413.65$.

%\subsection{Data analysis}

\begin{figure}[!t]
\centering
\resizebox{\columnwidth}{!}{\includegraphics{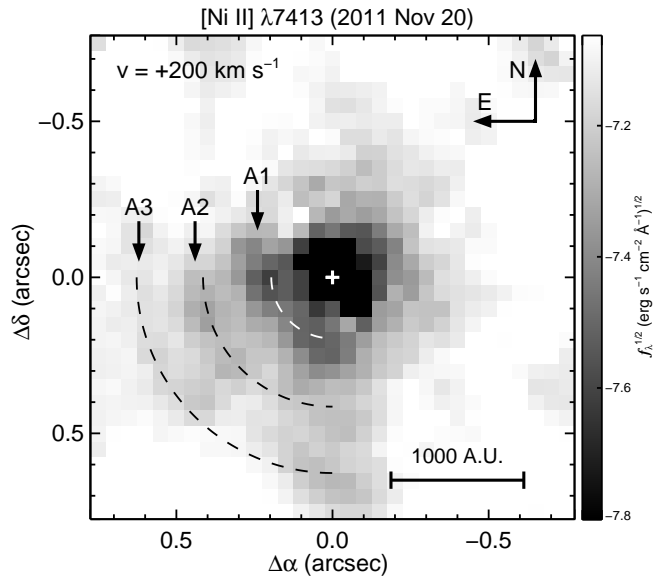}} \caption{Iso-velocity image of [\ion{Ni}{2}]~$\lambda7413$ at $+200$~km\,s$^{-1}$, obtained in 2011 November 20. The arcs discussed in this work are indicated by the arrows and were labeled A1, A2, and A3. The dashed lines show the expected position of three spherical shells traveling at a constant speed of $475$~km\,s$^{-1}$ for 4.5, 10, and 15.5 years, for the inner, middle, and outer arcs, respectively. To determine the expansion rate, we restricted our measurements to the eastern component of A2. \label{fig:refimg}}
\end{figure}

Visual examination of iso-velocity images in each data cube revealed multiple arcs in the low-ionization spectral lines used in this work. Sample images are presented in Figure~\ref{fig:panel} for the [\ion{Fe}{2}]~$\lambda4815$ transition. These arcs are visible between P.\,A. $90\arcdeg$ and $200\arcdeg$. We identified at least three sets of partial arcs, labelled A1, A2, and A3 in Figure~\ref{fig:refimg}. The two innermost arcs, A1 and A2, are conspicuously detected in all of the transitions from [\ion{Fe}{2}] and [\ion{Ni}{2}]. The outermost arc, A3, is very faint.

These arcs are not artifacts of the HST/STIS point spread function (PSF). The \textit{HST/STIS} PSF diffractive rings vary slowly with wavelength, are symmetric about the central core and have amplitudes substantially weaker than the partial rings apparent in the images \citep{2011SPIE.8127E..16K}. As shown in the discussion below, these features correlate with multiple forbidden line structures seen at different wavelengths, expanding away from the central source.

\begin{figure*}[!t]
\centering
\resizebox{2.09\columnwidth}{!}{\includegraphics{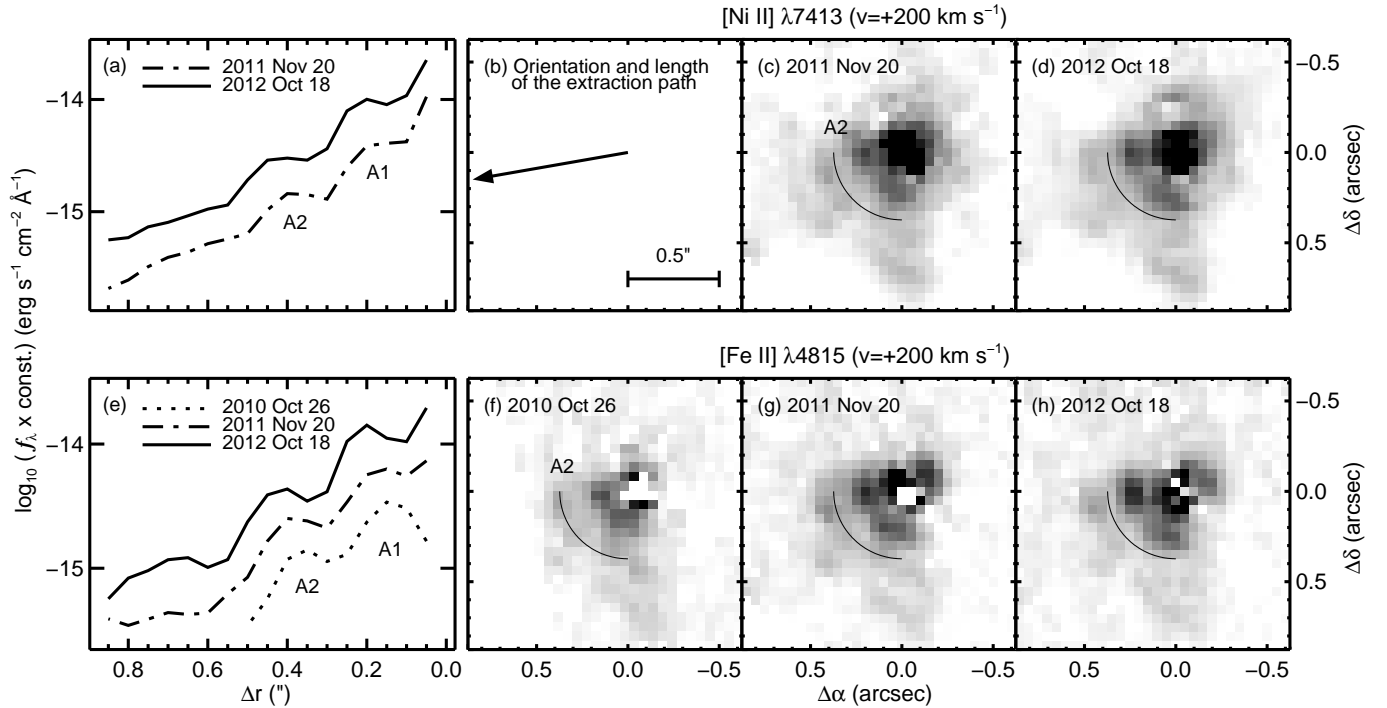}} \caption{Flux profiles and iso-velocity images of two forbidden lines: ($a$) and ($e$) are for [\ion{Ni}{2}]~$\lambda7413$ and [\ion{Fe}{2}]~$\lambda4815$, respectively, along the position indicated by the arrow in panel $(b)$, which has a length of approximately 0.9\arcsec~and ${\rm P.\,A.}=100\arcdeg$. The flux profiles were extracted from the iso-velocity images on the right. To illustrate the detected outward motion, in panels ($c$), ($d$), ($f$), ($g$), and ($h$) the curved solid line marks the position of A2 in 2010 October 26. For better visualization, the flux profiles were shifted in the vertical direction.\label{fig:profiles}}
\end{figure*}

Comparison of the same iso-velocity image for different epochs (2010 October 26, 2011 November 20, and 2012 October 18) revealed that A1, A2, and A3 move outward from the central source (Figure~\ref{fig:profiles}). Because A3 is faint and hard to see in all the images, and A1 is affected by uncertainties in the continuum subtraction method close to the central source, we concentrate our analysis on A2. We note, however, that all three arcs seem to expand at the same rate, but some extended parts become very distorted or diluted, which makes the determination of accurate positions a difficult task.

The position of the eastern portion of A2, relative to the central source, was measured using a radial flux profile cut along its direction (Figure~\ref{fig:profiles}). For each dataset (2010, 2011, and 2012), we extracted the flux profile between $0\arcsec<r<0.9\arcsec$ and $100\arcdeg<\rm{P.\,A.}<105\arcdeg$, where the emission from A2 is strongest.

We used three different methods for measuring the peak position in the radial flux profile curve. The first is a simple gaussian model, which is a good approximation when the flux profile of the shell is somewhat symmetric. The second method relies on the barycenter of the flux profile, and the third, developed by \citet{Blais:1986fi}, is based on the analysis of the derivative of the observed profile. The position of the peak is the median value of these methods.

A comprehensive study comparing several methods to determine peak positions with sub-pixel accuracy, including the three methods used here, was done by \citet{Fisher:1996de}. The reader is referred to that paper for further details. Based upon the individual measurement errors, we estimate an accuracy of 0.25 pixel (0.013\arcsec) for the position of the shell using the median value of the three methods mentioned before. The total error, however, may be larger than that because it is dominated by the dispersion in the measured position for various forbidden lines.

\section{Results}
\begin{figure*}[!t]
\centering
\mbox{\subfigure{\includegraphics[width=\columnwidth]{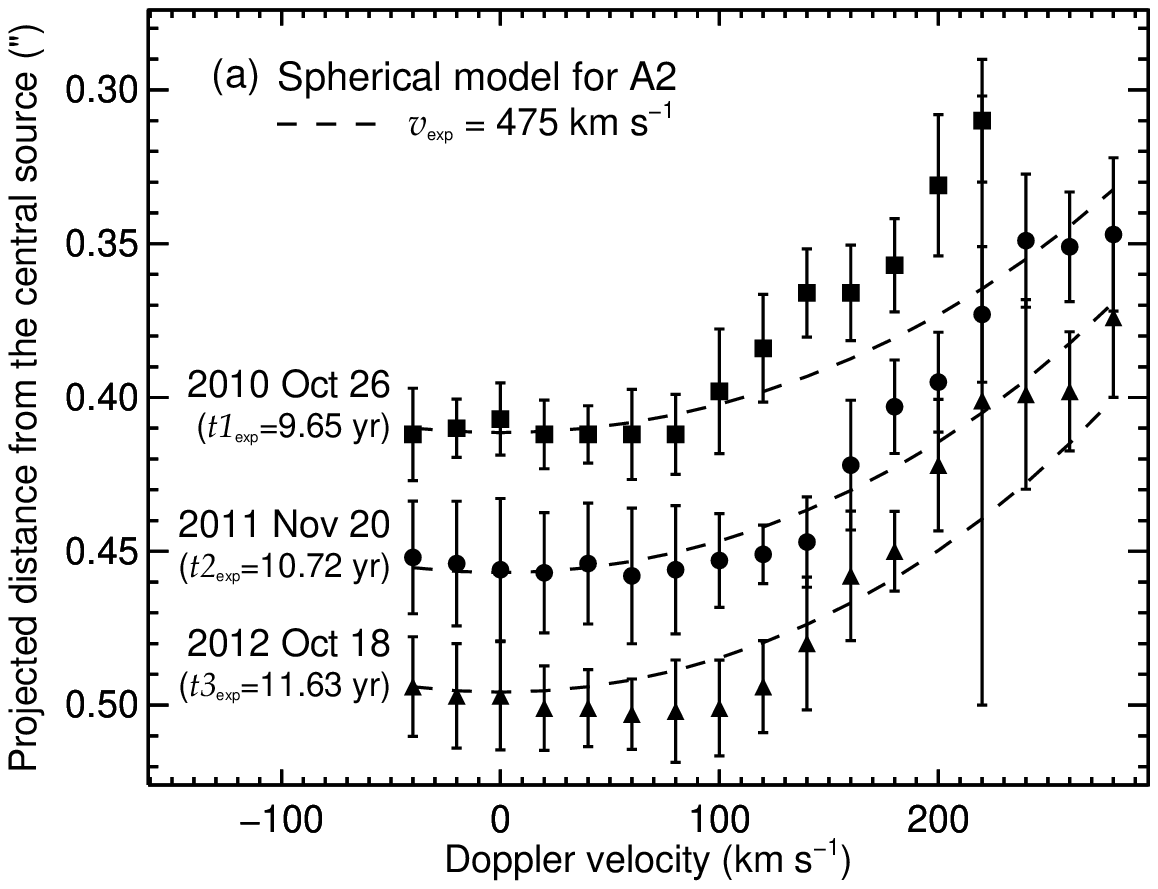}}\quad
      \subfigure{\includegraphics[width=\columnwidth]{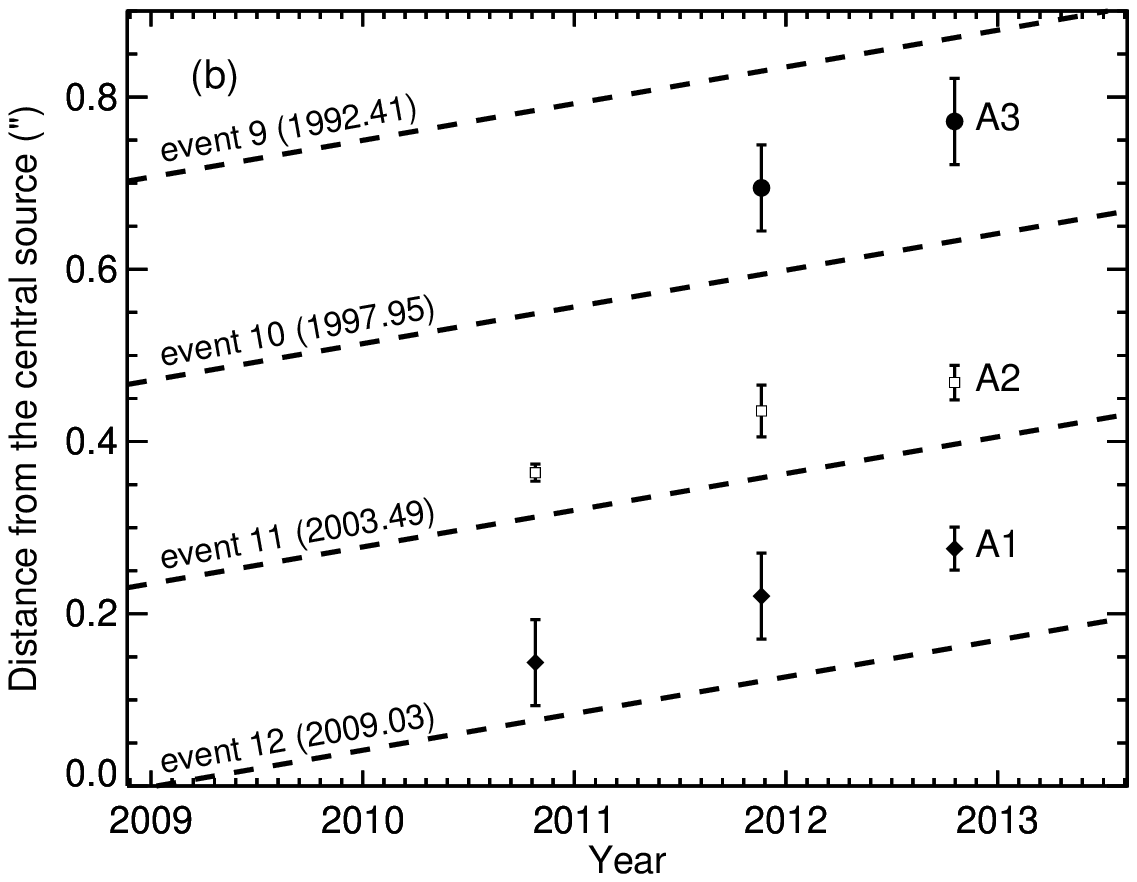}}} \caption{\btxt{($a$) Observed position of A2 as a function of Doppler velocity for the indicated epochs. The dashed lines show the projected position expected for a spherical shell expanding at $475$~km\,s$^{-1}$ during $t1_{\rm exp}$, $t2_{\rm exp}$, and $t3_{\rm exp}$. There is a clear departure from the spherical symmetry for red-shifted velocities $>+100$~km\,s$^{-1}$. ($b$) The dashed lines indicate the expected distance from the central source, as a function of time, of a spherical shell formed during periastron passage, and moving at $475$~km\,s$^{-1}$. The word 'event' corresponds to the periastron passage at the epoch indicated by the number in parenthesis. From this figure, A1 was formed in 2009.03, A2 in 2003.49, and A3 in 1997.95. In 2010.8, there is no data for A3 because this arc can only be detected in the [\ion{Ni}{2}] transition, and this transition was not included in the observations of that year.\label{fig:posvel}}}
\end{figure*}

We note that A2 is moving outwards with a proper motion $\mu=(0.05\pm0.01)$\arcsec\,yr$^{-1}$, on the plane of the sky (Figure~\ref{fig:posvel}(a)), a value suspiciously close to the pixel size. However, many other regions across the field of view are fixed in position from image to image, giving us confidence that the images are all spatially registered correctly. Furthermore, the motion of portions of A2 depends on the local Doppler velocity (\textit{cf.} Figure~\ref{fig:posvel}(a)). Hence, the detected motion is not caused by a systematic misalignment of the images.

The projected radius ($r_{\rm p}$), at a given Doppler velocity ($V_{\rm D}$), of a shell moving with space velocity $V$, during a time $t$, is given by
\begin{equation}
r_{\rm p}(t) = V \, t \, \sin\left(\arccos\left(\frac{V_{\rm D}}{V}\right)\right).
\label{eq:rp}
\end{equation}

Taking the difference of the projected radius of A2, at two epochs separated by a time interval $\Delta t = t_{2} - t_{1}$, leads to
\begin{equation}
r_{\rm p}(t_{2}) - r_{\rm p}(t_{1}) = V \, \Delta t \, \sin\left(\arccos\left(\frac{V_{\rm D}}{V}\right)\right).
\end{equation}

Setting the observations made in 2010 Oct. 26 as baseline, we have $\Delta t=1.07$ and $1.98$~yr, for the two subsequent datasets. Thus, we used the measured projected position for A2 in the range $-40 \leq V_D \leq +40$~km\,s$^{-1}$ (this is the small angle regime, where $V_D/V\ll 1$) to estimate the space velocity, which resulted in $V=(475\pm29)$~km\,s$^{-1}$. The age of A2, $t({\rm A2})$, in 2010, is thus obtained from $<r_{\rm p}({\rm A2})> / V = (9.65\pm0.55)$~yr, where $<r_{\rm p}({\rm A2})>=0.41\arcsec$ is the average projected distance of A2, relative to the central source, at that epoch.

\btxt{Assuming that A1 is moving at the same constant space velocity as A2}, then, for a given epoch, we have 
\begin{equation}
\frac{r_{\rm p}({\rm A1})}{r_{\rm p}({\rm A2})} = \frac{t({\rm A1})}{t({\rm A2})},
\label{eq:ratio}
\end{equation}
where $r_{\rm p}({\rm A1})$ is the average projected distance of A1 from the central source, and $t({\rm A1})$ is its age. In 2010 and 2011, A1 was too close to the central source \btxt{to be accurately measured}. Thus, we used the observations of 2012 to measure the position and then estimate the age. In 2012, we have $r_{\rm p}({\rm A1})\approx0.22\arcsec$, resulting in $t({\rm A1})=(5.98\pm0.92)$~yr. The relation described in equation~(\ref{eq:ratio}) also applies to A3, resulting in $t({\rm A3})=(17.13\pm0.95)$~yr, since it was observed at $r_{\rm p}({\rm A3})\approx0.6\arcsec$ in 2012. The rather large errors are dominated by uncertainties in the position due to the proximity to the central source for A1, and the weak emission of A3.

Correcting for the time interval between the 2012 and 2010 observations, and comparing the ages in 2012, we finally have $t({\rm A1})=5.98$, $t({\rm A2})=11.63$, and $t({\rm A3})=17.13$~yr. This means that each shell is created approximately 5.6~yr after the earlier one, showing that they are directly tied to the orbital period of the binary system \citep[$P=5.54$~yr;][]{2008MNRAS.384.1649D}.

\btxt{To test for the consistency of our results across the range of Doppler velocities where A2 is observed, we adopted a simplified model in which we consider A2 to be part of a spherical shell centered on $\eta$~Car, and allowed it to expand, at a constant space velocity of $475$~km\,s$^{-1}$, during the corresponding age of A2 at each epoch of observations ($t1_{\rm exp}$, $t2_{\rm exp}$, and $t3_{\rm exp}$ in Figure~\ref{fig:posvel}(a)).}

\btxt{At line of sight velocities near zero, a spherical model reproduces the observations very well, with standard deviations $<0.015\arcsec$. However, it fails to match the observed position of A2 at red-shifted velocities $>+100$~km\,s$^{-1}$; A2 moves slower than predicted. The final standard deviation for the spherical model is $0.05\arcsec$, in the range from $-40$ to $+220$~km\,s$^{-1}$.}

\btxt{Figure~\ref{fig:posvel}(b) shows the comparison between the observed and the expected motion of shells formed during various periastron passages, and moving at 475~km\,s$^{-1}$. Despite showing an offset between the expected and observed position at a given time, their slope are the same, which gives support to our derived space velocity and the assumption that the shells are expanding at the same rate.}

\section{Discussion and Conclusions}

We used the critical density for each forbidden line, calculated using the \href{http://www.chiantidatabase.org}{{\sc chianti}} atomic database \btxt{\citep{1997A&AS..125..149D,Landi:2013bb}}, to infer the typical local densities of the arcs. The outermost arc, A3, seen in the [\ion{Ni}{2}] transitions but weakly in the [\ion{Fe}{2}], must have a density in the range $0.8 - 4\times10^6$~cm$^{-3}$. The strong [\ion{Fe}{2}] emission from A1 and A2, indicates that their densities must be $>4\times10^6$~cm$^{-3}$.

\begin{figure*}[!t]
\centering
\resizebox{1.445\columnwidth}{!}{\includegraphics{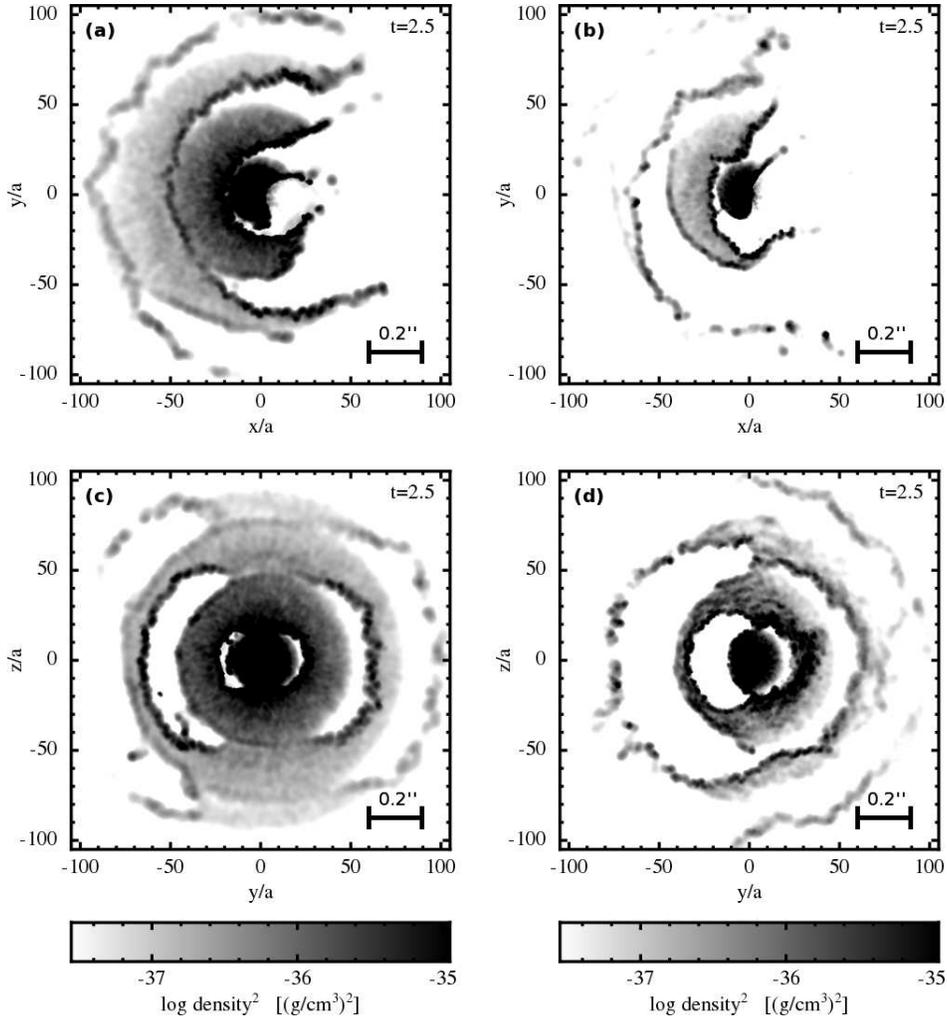}} \caption{3D SPH simulations showing the formation and geometry of high density regions due to the compression of the primary wind by the secondary wind, in two different regimes for the primary mass-loss rate: $8.5\times10^{-4}$~M$_\sun$\,yr$^{-1}$ (left) and $2.5\times10^{-4}$~M$_\sun$\,yr$^{-1}$ (right). \textit{Upper panels}: slices in the orbital plane (xy). \textit{Lower panels}: slices perpendicular to the orbital plane (yz). In the high mass-loss rate regime, there is a gap in the polar region that is not observed in the lower mass-loss rate regime. The lower and upper limits of the density scale correspond to the critical density values for [\ion{Ni}{2}] and [\ion{Fe}{2}], respectively.\label{sim}}
\end{figure*}

Assuming that A2 is half a spherical shell (\textit{cf.} Figure\,\ref{sim}) with average radius of $0.4\arcsec$ (projected distance from the central source, at zero Doppler velocity, in 2010) and thickness \btxt{$<0.05\arcsec$ (since we cannot spatially resolve the arcs)}, and using the critical density calculated before, we estimate a total mass of $2.6 \times 10^{-3}$~M$_\sun$, for solar metallicity.

Models for the primary star predict that its wind density should drop below $10^6$~cm$^{-3}$ for projected distances greater than 0.2\arcsec~\citep{2001ApJ...553..837H,2012MNRAS.423.1623G}. Since we detected arcs exceeding $10^6$~cm$^{-3}$ out to 0.6\arcsec, this is naturally explained by the primary wind being compressed -- a factor of $30$ -- by the secondary wind during periastron passage. This is also supported by the fact that the average age difference between the arcs is approximately $5.6$~yr, nicely matching the binary system's period.

The primary mass-loss rate required to produce a shell with the same mass and geometry as A2, over a time interval equal to the binary's period, is $9.3 \times 10^{-4}$~M$_\sun$\,yr$^{-1}$, \btxt{in agreement with the results from radiative transfer spectral modeling \citep{2001ApJ...553..837H,2006ApJ...642.1098H,2012MNRAS.423.1623G}}. We note, however, that our result must not be taken as a strict lower or upper limit, since the actual geometry and thickness of A2 may be different from what we assumed.

Simulations using 3D smoothed particle hydrodynamics (3D SPH) reveal compressed regions of the primary wind formed during the periastron passages and maintained for multiple cycles by the hot, low density, residual secondary wind (Figure~\ref{sim}). These simulations predict that these compressed regions accelerate from the primary wind terminal velocity \citep[$v_\infty=420$~km\,s$^{-1}$;][]{2012MNRAS.423.1623G} to about $460-470$~km\,s$^{-1}$ (T. Madura et al. 2013, in prep.), which is consistent with our derived expansion velocity of $475$~km\,s$^{-1}$ for the shells.

The departure from spherical symmetry for high red-shifted Doppler velocities suggests that the shells do not expand with the same velocity in all directions. This is explained by the fact that 3D SPH simulations show that material located in the orbital plane moves faster than material located off the plane. Thus, the observed Doppler velocity is a combination of velocities from different emitting regions along a specific viewing angle, which, in turn, depends upon the orientation of the shells on the sky, ultimately defined by the orbital orientation.

During periastron passages, SPH 3D simulations show that the secondary star creates a cavity deep within the primary wind, pushing a shell outwards. Since the low density secondary wind is moving at $\approx3000$~km\,s$^{-1}$, the interaction with the high density, relatively slow primary wind results in the formation of a high density, thin shell accelerated to velocities slightly higher than the terminal velocity of the primary wind. Hence, the geometry of the shell is strongly affected by the physical parameters of the binary system, and the projected appearance on the sky is defined by the inclination of the orbital plane.

The latitudinal extent of the shells is controlled by the asymptotic opening angle, $\theta_\infty$, of the wind-wind collision region, given by \citep{1992ApJ...389..635U}
\begin{equation}
\theta_\infty = \frac{2\pi}{3} \left( 1 - \frac{\eta^{2/5}}{4} \right) \eta^{1/3}\btxt{,}
\end{equation}
where $\eta=\dot{M}_{\rm B} \, v_{\infty B} / \dot{M}_{\rm A} \, v_{\infty A}$ is the ratio of the secondary to the primary wind momentum, $\dot{M}$ and $v_{\infty}$ are the mass-loss rate and terminal velocity of the wind, and the subscript A and B refer to the primary and secondary star, respectively.

Assuming fixed parameters for the secondary wind, a large primary mass loss rate decreases the bow shock opening angle $\theta_\infty$, so that the compressed primary wind material is trapped near the orbital plane (Figure\,\ref{sim}(c)); a low mass-loss rate increases $\theta_\infty$, and the compressed primary wind can extend well off the orbital plane (Figure\,\ref{sim}(d)). This means that the geometry of the circumstellar arcs is an excellent diagnostic of the primary mass-loss rate:  a lower mass loss rates produces a nearly continuous arc seen in projection around the star, while a higher mass loss rate produces a shorter, broken arc.

The existence of a discontinuity in the arcs that we observe, more pronounced in A1 as a lack of emission at P.\,A.$\approx135\arcdeg$, yields a lower limit to the mass-loss rate of the primary star. SPH simulations in 3D show that such broken rings are only apparent if the primary mass-loss rate is above $2.5\times10^{-4}$~M$_\sun$\,yr$^{-1}$, which suggests that, at least for the last event, the primary mass-loss rate must have exceeded this value.

Our results also constrain the orbital orientation, with periastron occuring between $200\arcdeg<\omega<270\arcdeg$, \textit{i.e.}, during periastron, the primary star is between the observer and the secondary star. The opposite configuration, where the secondary, at periastron, is between the observer and the primary star, would produce blue-shifted shells seen to the northwest of the central source, rather than the red-shifted features to the southeast that we observe.

Finally, we note that our results assume a constant shell space velocity. \btxt{However, as can be seen in Figure\,\ref{fig:posvel}(b), there is an offset of 0.07\arcsec~between the expected and the observed distance from the central source as a function of time. We know, from the comparison between the derived space velocity (475~km\,s$^{-1}$) and the terminal velocity of the primary wind (420~km\,s$^{-1}$), that this offset is due to the acceleration that these shells experienced at some point during the early stages of their formation. Thus, the ages derived in this letter are overestimated, \textit{i.e} the shells are younger. Nevertheless, the difference of 5.6~yr between them remains the same, validating their connection with the orbital period.}

\acknowledgments
\btxt{We are grateful to an anonymous referee for comments and suggestions that helped to improve the quality and presentation of this work.} M.\,T. is supported by CNPq/MCT-Brazil through grant 201978/2012-1. T. I. M. was supported by an appointment to the NASA Postdoctoral Program at the Goddard Space Flight Center, administered by Oak Ridge Associated Universities through a contract with NASA. We thank Don Lindler for his consistent help with the {\sc idl} procedures necessary for production of the data cubes. M.\,T. would like to thank David Fanning for making his \href{http://www.idlcoyote.com/index.html}{{\sc idl} Coyote library} publicly available. CHIANTI is a collaborative project involving George Mason University, the University of Michigan (USA) and the University of Cambridge (UK). This research has made extensive use of NASA's Astrophysics Data System and \href{http://idlastro.gsfc.nasa.gov}{{\sc idl} Astronomy User's Library}.

{\it Facilities:} \facility{HST (STIS)}.


\begin{thebibliography}{}
\expandafter\ifx\csname natexlab\endcsname\relax\def\natexlab#1{#1}\fi

\bibitem[{Blais \& Rioux(1986)}]{Blais:1986fi}
Blais, F., \& Rioux, M. 1986, Signal Processing, 11, 145

\bibitem[{Bostroem \& Proffitt(2011)}]{Bostroem:2011wg}
Bostroem, K.~A., \& Proffitt, C. 2011, STIS Data Handbook, HST Data Handbooks,
  -1

\bibitem[{Damineli {et~al.}(2008)Damineli, Hillier, Corcoran, Stahl,
  Levenhagen, Leister, Groh, Teodoro, Albacete~Colombo, Gonzalez, Arias,
  Levato, Grosso, Morrell, Gamen, Wallerstein, \&
  Niemela}]{2008MNRAS.384.1649D}
Damineli, A., Hillier, D.~J., Corcoran, M.~F., {et~al.} 2008, Monthly Notices
  of the Royal Astronomical Society, 384, 1649

\bibitem[{Davidson \& Humphreys(1997)}]{1997ARA&A..35....1D}
Davidson, K., \& Humphreys, R.~M. 1997, Annual Review of Astron and Astrophys,
  35, 1

\bibitem[{Dere {et~al.}(1997)Dere, Landi, Mason, Monsignori~Fossi, \&
  Young}]{1997A&AS..125..149D}
Dere, K.~P., Landi, E., Mason, H.~E., Monsignori~Fossi, B.~C., \& Young, P.~R.
  1997, A {\&} A Supplement series, 125, 149

\bibitem[{Fisher \& Naidu(1996)}]{Fisher:1996de}
Fisher, R.~B., \& Naidu, D.~K. 1996, in link.springer.com (Berlin, Heidelberg:
  Springer Berlin Heidelberg), 385--404

\bibitem[{Groh {et~al.}(2012)Groh, Hillier, Madura, \&
  Weigelt}]{2012MNRAS.423.1623G}
Groh, J.~H., Hillier, D.~J., Madura, T.~I., \& Weigelt, G. 2012, Monthly
  Notices of the Royal Astronomical Society, 423, 1623

\bibitem[{Gull {et~al.}(2011)Gull, Madura, Groh, \&
  Corcoran}]{2011ApJ...743L...3G}
Gull, T.~R., Madura, T.~I., Groh, J.~H., \& Corcoran, M.~F. 2011, The
  Astrophysical Journal Letters, 743, L3

\bibitem[{Gull {et~al.}(2009)Gull, Nielsen, Corcoran, Madura, Owocki, Russell,
  Hillier, Hamaguchi, Kober, Weis, Stahl, \& Okazaki}]{2009MNRAS.396.1308G}
Gull, T.~R., Nielsen, K.~E., Corcoran, M.~F., {et~al.} 2009, Monthly Notices of
  the Royal Astronomical Society, 396, 1308

\bibitem[{Hillier {et~al.}(2001)Hillier, Davidson, Ishibashi, \&
  Gull}]{2001ApJ...553..837H}
Hillier, D.~J., Davidson, K., Ishibashi, K., \& Gull, T. 2001, The
  Astrophysical Journal, 553, 837

\bibitem[{Hillier {et~al.}(2006)Hillier, Gull, Nielsen, Sonneborn, Iping,
  Smith, Corcoran, Damineli, Hamann, Martin, \& Weis}]{2006ApJ...642.1098H}
Hillier, D.~J., Gull, T., Nielsen, K., {et~al.} 2006, The Astrophysical
  Journal, 642, 1098

\bibitem[{Krist {et~al.}(2011)Krist, Hook, \& Stoehr}]{2011SPIE.8127E..16K}
Krist, J.~E., Hook, R.~N., \& Stoehr, F. 2011, Optical Modeling and Performance
  Predictions V. Edited by Kahan, 8127, 16

\bibitem[{Landi {et~al.}(2013)Landi, Young, Dere, Del~Zanna, \&
  Mason}]{Landi:2013bb}
Landi, E., Young, P.~R., Dere, K.~P., Del~Zanna, G., \& Mason, H.~E. 2013,
  Astrophysical Journal, 763, 86

\bibitem[{Madura {et~al.}(2012)Madura, Gull, Owocki, Groh, Okazaki, \&
  Russell}]{2012MNRAS.420.2064M}
Madura, T.~I., Gull, T.~R., Owocki, S.~P., {et~al.} 2012, Monthly Notices of
  the Royal Astronomical Society, 420, 2064

\bibitem[{Mehner {et~al.}(2010)Mehner, Davidson, Ferland, \&
  Humphreys}]{2010ApJ...710..729M}
Mehner, A., Davidson, K., Ferland, G.~J., \& Humphreys, R.~M. 2010, The
  Astrophysical Journal, 710, 729

\bibitem[{Parkin {et~al.}(2011)Parkin, Pittard, Corcoran, \&
  Hamaguchi}]{2011ApJ...726..105P}
Parkin, E.~R., Pittard, J.~M., Corcoran, M.~F., \& Hamaguchi, K. 2011, The
  Astrophysical Journal, 726, 105

\bibitem[{Parkin {et~al.}(2009)Parkin, Pittard, Corcoran, Hamaguchi, \&
  Stevens}]{2009MNRAS.394.1758P}
Parkin, E.~R., Pittard, J.~M., Corcoran, M.~F., Hamaguchi, K., \& Stevens,
  I.~R. 2009, Monthly Notices of the Royal Astronomical Society, 394, 1758

\bibitem[{Pittard \& Corcoran(2002)}]{2002A&A...383..636P}
Pittard, J.~M., \& Corcoran, M.~F. 2002, Astronomy and Astrophysics, 383, 636

\bibitem[{Smith(2006)}]{2006ApJ...644.1151S}
Smith, N. 2006, The Astrophysical Journal, 644, 1151

\bibitem[{Teodoro {et~al.}(2008)Teodoro, Damineli, Sharp, Groh, \&
  Barbosa}]{2008MNRAS.387..564T}
Teodoro, M., Damineli, A., Sharp, R.~G., Groh, J.~H., \& Barbosa, C.~L. 2008,
  Monthly Notices of the Royal Astronomical Society, 387, 564

\bibitem[{Usov(1992)}]{1992ApJ...389..635U}
Usov, V.~V. 1992, Astrophysical Journal, 389, 635

\bibitem[{Verner {et~al.}(2005)Verner, Bruhweiler, \&
  Gull}]{2005ApJ...624..973V}
Verner, E., Bruhweiler, F., \& Gull, T. 2005, The Astrophysical Journal, 624,
  973

\bibitem[{Zethson {et~al.}(2012)Zethson, Johansson, Hartman, \&
  Gull}]{2012A&A...540A.133Z}
Zethson, T., Johansson, S., Hartman, H., \& Gull, T.~R. 2012, Astronomy and
  Astrophysics, 540, 133

\end{thebibliography}
\end{document}